\documentclass[]{article}

\usepackage{multirow, amsmath, amsthm, graphicx, amssymb, amsfonts, latexsym, enumerate, amscd, xcolor, float, caption, url}
\usepackage{listings}
\usepackage{geometry}
\RequirePackage[numbers]{natbib}

\begin{document}
\title{Low-rank Variational Bayes correction to the Laplace method}

\author{Janet van Niekerk (janet.vanNiekerk@kaust.edu.sa )\\
	AND \\
	H\aa vard Rue (haavard.rue@kaust.edu.sa )\\
	CEMSE Division\\
	King Abdullah University of Science and Technology\\
	Kingdom of Saudi Arabia}

\maketitle

\begin{abstract}
	Approximate inference methods like the Laplace method, Laplace approximations and
	variational methods, amongst others, are popular methods when
	exact inference is not feasible due to the complexity of the model
	or the abundance of data. In this paper we propose a hybrid
	approximate method called Low-Rank Variational Bayes correction
	(VBC), that uses the Laplace method and subsequently a Variational
	Bayes correction in a lower dimension, to the joint posterior mean. The cost is essentially
	that of the Laplace method which ensures scalability of the
	method, in both model complexity and data size. 
	Models with fixed
	and unknown hyperparameters are considered, for simulated and real examples, for small and large datasets.
\end{abstract}

\section{Introduction}
Bayesian methods involve a prior belief about a model and learning
from the data to arrive at a new belief, which is termed the posterior
belief. Mathematically, the posterior belief can be derived from the
prior belief and the empirical evidence presented by the data using
Bayes' rule. In this way Bayesian analysis is a
natural statistical machine learning method (see \citet{theodoridis2015, 
	chen2016, polson2017, rehman2019, sambasivan2020, vehtari2020, 
	moss2021, richardson2021} amongst many others), and especially 
powerful
for small datasets, missing data or complex models. Suppose we observe data $\pmb y$ for which we formulate a data generating model, $\mathcal F$. Suppose we have unknown parameters $\pmb\psi$ in $\mathcal F$ for which we can define priors. Then we want to find the posterior inference of $\pmb\psi$, denoted as $\pi(\pmb\psi|\pmb y)$. The question of how to calculate $\pi(\pmb\psi|\pmb y)$ now arises.  

From a computational viewpoint, various approaches have been proposed
to perform Bayesian analysis, mainly exact (analytical or
sampling-based) or approximate inferential approaches.  
Sampling-based
methods like Markov Chain Monte Carlo (MCMC) sampling with its
extensions (see \citet{metropolis1953, geman1984, casella1992, 
	andrieu2003}, amongst others) gained popularity in the 1990's 
but suffers from slow speed
and convergence issues exacerbated by large data and/or complex
models. Hamiltonian Monte Carlo (HMC) methods, as implemented in the
STAN software, are showing promise for more efficient sampling-based inference. 
Partly motivated by the inefficency of sampling-based methods, approximate methods were 
developed to \emph{approximate} the posterior density as a more efficient means of inference. Not all
approximate methods are equally accurate or efficient. Some approximate 
methods are essentially sampling-based like the Monte Carlo Adjusted 
Langevin Algorithm (MALA) \citep{rossky1978, roberts1996}, pseudo-
marginal MCMC \citep{andrieu2009} and Approximate Bayesian Computation
(ABC) \citep{beaumont2002, tavare1997}, and thus still slow. Asymptotic approximate 
methods are not sampling-based and propose a specific form of the 
posterior like the Laplace method \citep{van2000, laplace1986memoir, 
	tierney1989fully} and the Integrated Nested Laplace Approximation 
(INLA)  \citep{rue2009, bakka2018, van2019new} for Latent Gaussian 
models. Optimization-based approximate methods like Variational Bayes
(VB) \citep{attias1999, jordan1999, blei2017, hoffman2013stochastic}, 
Expectation
Propagation (EP) \citep{opper2000gaussian, minka2001, dehaene2016} 
and discrete distributions approximations by \citet{liu2016, nitanda2017} 
are also popular.  

The Laplace method using a second-order series expansion around the mode \citep{tierney1989fully, laplace1986memoir}, is a common approach to calculate an approximate Gaussian density to an unknown function. Its popularity can be assigned to its simplicity and low computational cost. It is mostly used to approximate marginal likelihoods, or marginal posteriors in a Bayesian framework. The Hessian of the unknown function, evaluated at the mode provides a quantification of the uncertainty, and under some regularity assumptions is a consistent estimator (\cite{newey1994}). From the Laplace method we can thus approximate $\pi(\pmb\psi|\pmb y)$ as
\begin{equation*}
	\tilde\pi(\pmb\psi|\pmb y) \propto \exp\left(-\frac{1}{2}(\pmb\psi - \pmb\mu)^\top\pmb Q (\pmb\psi - \pmb\mu) \right),
\end{equation*}
such that $-\pmb Q$ is the Hessian matrix of $\log\pi(\pmb\psi|\pmb y)$ evaluated at $\pmb\mu$, the mode of $\log\pi(\pmb\psi|\pmb y)$.  

When the function is not uni-modal or exhibit heavy-tail behavior, the Laplace method does not provide an accurate approximation to the function and other families besides the Gaussian could be considered. The Gaussian assumption is often too strict for marginal posteriors and more flexible families should be considered that would allow for some skewness or heavier tails. Variational Bayes (VB) methods are based on the optimization of a certain objective function (variational energy resulting in the evidence lower bound) for a specific family of distributions. As such, any family can be considered and the Gaussian assumption of the marginal posteriors is not needed.  

Suppose we posit that the approximate posterior of $\pmb\psi$ comes from family $\mathcal G$, with members $g$, then the VB approximation of $\pi(\pmb\psi|\pmb y)$  is $\tilde\pi(\pmb\psi|\pmb y) = g(\pmb\psi)$, such that 
\begin{equation}
	\tilde\pi(\pmb\psi|\pmb y) = \arg\min_{g\in \mathcal G} \text{KLD}(g(\pmb\psi)||\pi(\pmb\psi|\pmb y)),\label{eq:vb}
\end{equation}
where $ \text{KLD}(g||h)$ is the Kullback-Leibler divergence from probability distribution $g$ to probability distribution $h$.
Now since $\pi(\pmb\psi|\pmb y)$ is unknown, it is shown that the minimizer is also the maximizer of the evidence lower bound (ELBO), such that 
\begin{equation}
	\tilde\pi(\pmb\psi|\pmb y) = \arg\max_{g\in \mathcal G} \text{E}_{g}(\log g(\pmb\psi) - \log \pi(\pmb\psi,\pmb y)).\label{eq:elbo}
\end{equation}
For a specific choice of $\mathcal G$, specialized optimization techniques can be developed and applied. Some works for $\mathcal G$ being the Gaussian family are the Gaussian flow or
Gaussian particle flow techniques \citep{galy2021}, Stein Variational Gradient
Descent \citep{zhuo2018, korba2020, lu2019}, recursive VGI
\citep{lambert2020} and exactly sparse VGA \citep{barfoot2020},
amongst others. If the selected family includes the true posterior, then the variational Bayes approximation could recover the true posterior. Variational frameworks, however, are known to suffer from severe underestimation of the uncertainty of the point estimate, due to the non-availability of a consistent estimator of the variance of the variational estimate for an often simple form of $\mathcal{G}$ (see for example the Appendix of \citet{rue2009} for more details). This underestimation will produce poor credible intervals and could result in incorrect decision-making. Furthermore, the parameters of the chosen family in \eqref{eq:elbo} should all be estimated and the optimization problem should be solved in the dimension of the parameter space, which can be very large in for example spatial models. Even if scalable (in some sense) approaches are proposed to optimize \eqref{eq:elbo}, all unknown parameters will have to be solved for.  

We consider the case where a Gaussian approximation is opted for (not necessarily for the marginals). Our assumption is that the unknown density is uni-modal, and thus the Hessian matrix provides a reasonable estimate of the curvature at the mode.
In this paper we present a novel approach to approximate an unknown density function with a Gaussian density function, that provides reasonable first and second order properties. We achieve this by employing the Laplace method, and then we formulate a low-rank variational correction to the mode of this Gaussian approximation. The variational correction to the Laplace method's mode, is defined in dimension $p$, that is much smaller than the latent field dimension $m$. This is possible since we learn the graph of connectedness from the Laplace method, and we use that to propagate any change in the lower dimension to all elements in the higher dimensional latent field.  

Although our proposal can be used in various ways, we show the impact it has in the Bayesian inference of latent Gaussian models by applying the proposal \emph{not} to the latent marginal posteriors, but to the latent \emph{conditional posteriors}, since the latent conditional posteriors are in fact more Gaussian-like as shown by \citet{rue2009}. This provides an accurate and very efficient approximate Bayesian inference tool for latent Gaussian models that include generalized additive mixed models, spatial models, temporal models, lifetime analysis models and many more.

\section{Proposal}\label{sec:proposal}
Based on data $\pmb y$ of size $n$, and unknown latent set $\pmb\psi\in \mathbb{R}^m$, we formulate a data generating model, $\pi(\pmb y|\pmb\psi)$ that depends on $\pmb\psi$, such that the data is conditionally independent given $\pmb\psi$. The goal is to infer $\pmb\psi$ based on the data $\pmb y$ and elective external information (prior information) $\pi(\pmb\psi)$. The joint density then is $\pi(\pmb\psi,\pmb y)$. From this we can use Bayes' theorem to formulate the posterior density of $\pmb\psi$ as
\begin{equation*}
	\pi(\pmb\psi|\pmb y) = \frac{\pi(\pmb y|\pmb\psi)\pi(\pmb\psi)}{\pi(\pmb y)}.
\end{equation*}
The Gaussian approximation of $\pi(\pmb\psi|\pmb y)$ from the Laplace method is then derived from
\begin{equation*}
	\ln(\pi(\pmb\psi|\pmb y)) = 	\ln(\pi(\pmb\psi_0|\pmb y) - \frac{1}{2}(\pmb\psi - \pmb\psi_0)^\top \pmb H|_{\pmb\psi = \pmb\psi_0}(\pmb\psi - \pmb\psi_0) + \text{higher order terms},
\end{equation*}
where $\pmb\psi_0$ is the mode of $\ln(\pi(\pmb\psi|\pmb y))$ and $\pmb H$ is the negative Hessian matrix. Then
\begin{equation}
	\tilde{\pi}(\pmb\psi|\pmb y) \propto \exp\left(-\frac{1}{2}(\pmb\psi -\pmb\psi_0)^\top \pmb H|_{\pmb\psi = \pmb\psi_0}(\pmb\psi - \pmb\psi_0)\right),\label{eq:ga}
\end{equation}
so that $\pmb\psi|\pmb y \dot\sim N(\pmb\psi_0, \pmb H^{-1}|_{\pmb\psi = \pmb\psi_0})$ (approximately distributed as). To find the mode we solve for $ \pmb\psi_0$ in
the linear system
\begin{equation}
	\pmb H|_{\pmb\psi = \pmb\psi_0} \pmb\psi_0 = \pmb\gamma|_{\pmb\psi = \pmb\psi_0} + \pmb H|_{\pmb\psi = \pmb\psi_0} \pmb\psi_0,
	\label{eq:hmu=b}
\end{equation}
where $\pmb\gamma|_{\pmb\psi = \pmb\psi_0}$ is the gradient of $\ln(\pi(\pmb\psi|\pmb y))$ evaluated at $\pmb\psi = \pmb\psi_0$. Now let $\pmb Q_0 = \pmb H|_{\pmb\psi = \pmb\psi_0}$ and $\pmb b_0 = \pmb\gamma|_{\pmb\psi = \pmb\psi_0} + \pmb H|_{\pmb\psi = \pmb\psi_0} \pmb\psi_0$, then the system can be written as 
\begin{equation}
	\pmb Q_0\pmb\psi_0 = \pmb b_0.
	\label{eq:qmu=b}
\end{equation}
The precision matrix $\pmb Q_0$, relates information about the conditional dependence amongst the elements in $\pmb\psi$. Since the approximation in \eqref{eq:ga} is an approximation to the \emph{joint} posterior, we still need to calculate the marginal posteriors. It is well-known that the marginal posteriors based on a joint Gaussian distribution can be computed as univariate Gaussian densities based on the elements of the joint mean and the diagonal elements of the inverse precision matrix, making the multivariate Gaussian assumption attractive. 

We want to correct the mean of the Gaussian approximation to have a more accurate mean that is not necessarily the MAP (maximum a posteriori) estimator. As such we propose an updated mean,
\begin{equation}
	\pmb\psi_1 = \pmb\psi_0 + \pmb\delta,
\end{equation}
where $\pmb\delta$ can be viewed as corrections to the MAP estimator, such that the approximate posterior of $\pmb\psi$ is then
\begin{equation*}
	\tilde\pi(\pmb\psi|\pmb y) = (2\pi)^{-m/2}|\pmb Q_0|^{1/2}\exp\left(-\frac{1}{2}(\pmb\psi - \pmb\psi_1)^\top\pmb Q_0(\pmb\psi - \pmb\psi_1)\right),\quad \pmb\psi\in\mathbb{R}^m
\end{equation*}
where $\pmb\psi_1 = \pmb\psi_0 + \pmb\delta$.  Now the question arises: how can we estimate $\pmb\delta$ in a fast and accurate way?  

Since the dimension of $\pmb\psi$ is $m$, we would need to find $m$ values that produce a more accurate joint posterior mean. If the model is complex or contains many random effects, this dimension can be very large. For efficiency, we can use a variational framework since we only want to find a more accurate mean, while fixing the precision matrix based on the calculated Hessian. The ELBO for this problem is
\begin{equation}
	\text{E}_{\pmb\psi\sim N(\pmb\psi_0 + \pmb\delta, \pmb Q_0^{-1})}(\log \phi(\pmb\psi|\pmb\psi_0 + \pmb\delta, \pmb Q_0^{-1}) - \log \pi(\pmb\psi,\pmb y)),\label{eq:elbo_delta}
\end{equation}
where $\phi(.)$ is the Gaussian density function. This optimization can be done in various ways using many specialized techniques proposed in literature on maximizing ELBO's.  

Rather than working with the ELBO, we revert back to the fundamental idea of Variational Bayes as introduced by \citet{zellner1988optimal} (for more details see the Appendix) and more recently posed as an optimization view of Bayes' rule by \citet{knoblauch2022}. Based on the available information from the prior of the unknown parameters $\pi(\pmb\psi)$ and the conditional likelihood of the data $\pi(\pmb y|\pmb\psi)$, we can derive two outputs: the marginal likelihood of the data $\pi(\pmb y)$ and the posterior of the unknown parameters $\pi(\pmb\psi|\pmb y)$. If we want to use the input information optimally, then we find the approximate posterior $\tilde\pi(\pmb\psi|\pmb y)$, such that
\begin{equation}
	\tilde\pi(\pmb\psi|\pmb y) = \arg\min_{g\in  \mathcal{G}} E_{\pmb\psi}[-\log\pi(\pmb y|\pmb\psi)] + \text{KLD}(g||\pi(\pmb\psi)).
	\label{eq:zellnervb}
\end{equation}
In the work of \citet{zellner1988optimal}, it was shown that this variational framework produces the true posterior, from the appropriate family, as calculated from Bayes' theorem and thus implying that Bayes' theorem is an optimal rule for processing of information. Note that \eqref{eq:zellnervb} does not contain the unknown true posterior $\pi(\pmb\psi|\pmb y)$ as in \eqref{eq:vb}, and can be directly optimized. Thus to use \eqref{eq:zellnervb} for the mean correction, we need to calculate
\begin{equation}
	\tilde{\pmb\delta} =\arg\min_{\pmb\delta} E_{\pmb\psi\sim N(\pmb\psi_0 + \pmb\delta, \pmb Q_0^{-1})}[-\log\pi(\pmb y|\pmb\psi)] + \text{KLD}\left(\phi(\pmb\psi|\pmb\psi_0 + \pmb\delta,\pmb Q_0^{-1})||\pi(\pmb\psi)\right)
	\label{eq:vb_delta}
\end{equation}
Whichever method is used to solve \eqref{eq:vb_delta}, the optimization is over an $m$-dimensional vector, thus the computational and memory cost will be based on $m$, which can be large.

\subsection{Low-rank variational correction}\label{sec:vbc_lowrank}
Rather than an explicit correction to the MAP, we propose an implicit correction, by explicitly correcting the estimated gradient such that the improved posterior mean $\pmb\psi_1$, satisfies the new linear system,
\begin{equation}
	\pmb Q_0\pmb\psi_1 = \pmb b_0 +\pmb\lambda = \pmb b_1.
\end{equation}
Now, if $\pmb\lambda\in \mathbb{R}^m$ then we would not gain any computational advantage over the proposal in \eqref{eq:vb_delta}, but because of the linear system, a change to any element in $\pmb b_1$ will propagate changes to all the elements in $\pmb\psi_1$. For a non-zero value of the $j^{\text{th}}$ element of $\pmb\lambda$ i.e. $\lambda_j \neq 0$, the change this value causes to the $i^{\text{th}}$ element of $\pmb\psi_1$, $\psi_{1,i}$ is  
\begin{equation}
	\frac{\partial\psi_{1,i}}{\partial \lambda_j} = \frac{\partial\psi_{1,i}}{\partial b_{1,j}} \frac{\partial b_{1,j}}{\partial \lambda_j} =  Q^{ij}_{0}\lambda_j
\end{equation}
where $ Q^{ij}$ denotes the element in the $i^{\text{th}}$ row and $j^{\text{th}}$ column of the inverse of $\pmb Q$. Thus, in vector notation, 
\begin{equation}
	\frac{\partial\pmb\psi_{1}}{\partial \lambda_j} = \pmb Q^{. j}_{0}\lambda_j
\end{equation}
where $\pmb Q^{. j}$ denotes the $j^{\text{th}}$ column of the inverse of $\pmb Q$. This enables us to propose a low-rank Variational Bayes correction (VBC) since the dimension of $\pmb\lambda$ is $p$, which can be much smaller than $m$ and $n$, and $p$ does not have to grow with $m$ or $n$.  

Suppose we have a set of indices $i\in I\subset\{1,2,...,m\}$ for which we want to correct $b_{0,i}$, then we extract the relevant columns of $\pmb Q_0^{-1}$ and denote it by $\pmb Q_I^{-1}$. The improved mean is thus
\begin{equation*}
	\pmb\psi_1 = \pmb\psi_0 + \pmb Q_I^{-1}\pmb\lambda.
\end{equation*}
Now we can optimize \eqref{eq:vb_delta}, but for $\pmb\lambda$ in dimension $p$ instead of $\pmb\delta$ in dimension $m$ as follows
\begin{equation}
	\tilde{\pmb\lambda} = \arg\min_{\pmb\lambda} E_{\pmb\psi\sim N(\pmb\psi_0 + \pmb Q^{-1}_I\pmb\lambda, \pmb Q_0^{-1})}[-\log\pi(\pmb y|\pmb\psi)] + \text{KLD}\left(\phi(\pmb\psi|\pmb\psi_0 + \pmb Q^{-1}_I\pmb\lambda,\pmb Q_0^{-1})||\pi(\pmb\psi)\right).\label{eq:vb_lambda}
\end{equation}
This proposal allows us to correct an $m$-dimensional MAP estimator with a rank $p$ update with $p\ll m$, resulting in a computational cost of about $O(mp^2)$, since we do not need to calculate the entire inverse of $\pmb Q_0$ but only the selected elements based on $I$. Moreover, from \citet{zellner1988optimal}, this optimization is optimally information efficient and converges to the true posterior when the true family is selected. We illustrate this convergence using simulated and real examples, and we compare the posterior from a Gaussian approximation with the VB correction, to the posterior from MCMC samples in Section \ref{sec:poisson_example_gen}.  

Our proposal to approximate the joint posterior can be summarized as follows:
\begin{enumerate}
	\item Calculate the gradient $\pmb\gamma$, and the negative Hessian matrix $\pmb H$, of $\log\pi(\pmb\psi|\pmb y)$.
	\item Find the MAP estimator by solving for $\pmb\psi_0$ such that 
	\begin{equation*}
		\pmb H|_{\pmb\psi = \pmb\psi_0} \pmb\psi_0 = \pmb\gamma|_{\pmb\psi = \pmb\psi_0} + \pmb H|_{\pmb\psi = \pmb\psi_0} \pmb\psi_0,
	\end{equation*}
	and define $\pmb Q_0 = \pmb H|_{\pmb\psi = \pmb\psi_0}$ and $\pmb b_0 = \pmb\gamma|_{\pmb\psi = \pmb\psi_0} + \pmb H|_{\pmb\psi = \pmb\psi_0} \pmb\psi_0$.
	\item Decide on the set of indices for correction, $I$, construct the $p\times m$ matrix $\pmb Q^{-1}_I$ from the columns of the inverse of $\pmb Q_0$, $\pmb Q_0^{-1}$, and solve for $\pmb\lambda$ such that 
	\begin{equation*}
		\tilde{\pmb\lambda} = \arg\min_{\pmb\lambda} E_{\pmb\psi\sim N(\pmb\psi_0 + \pmb Q^{-1}_I\pmb\lambda, \pmb Q_0^{-1})}[-\log\pi(\pmb y|\pmb\psi)] + \text{KLD}\left(\phi(\pmb\psi|\pmb\psi_0 + \pmb Q^{-1}_I\pmb\lambda,\pmb Q_0^{-1})||\pi(\pmb\psi)\right).
	\end{equation*}
	\item The approximate posterior of $\pmb\psi$ is Gaussian with mean $\pmb\psi_1 = \pmb\psi_0 + \pmb Q^{-1}_I\tilde{\pmb\lambda}$ and precision matrix $\pmb Q_0$.
\end{enumerate}
Now we consider the choice of the index set $I$. 
Since a change in any one element of $\pmb b_1$ is propagated to the posterior mean of the entire latent field, similar choices of the index set $I$, will result in a similar improved joint posterior mean, since the proposal is based on an improved joint Gaussian approximation for the entire field. We are thus not solving for element-wise corrections, and from the work of \citet{zellner1988optimal} we are assured of a joint improvement. 
From our experience, we want to mainly correct the Gaussian approximation for those elements in $\pmb\psi$ that are most influential and connected to many datapoints. 
Hence,  
we explicitly correct
the posterior means of the fixed effects, and those random effects that are connected to many datapoints (short length random effects).
We return to this in Sections \ref{sec:poisson_example_gen}, \ref{sec:sim_inla} and \ref{sec:real_examples}.  

Even though the proposal looks basic and simple, various computational details are intricate and complicated to ensure a low computational cost while maintaining accuracy. Some of these details are presented in the next section.

\subsection{Computational aspects}
In this section we focus on computational aspects regarding the proposed variation Bayes correction to the Laplace method. The gradient and Hessian matrix, can be calculated numerically and we can use various gradient descent or Newton-Raphson type algorithms. In our approach we use the smart gradient proposed by \citet{fattah2022}. The efficient calculation of the expected log-likelihood in \eqref{eq:vb_lambda} requires some attention and we present our approach in this section.
\subsubsection{Smart gradient}
Numerical gradients are important in various optimization techniques (as in our proposal) such as stochastic gradient descent, trust region and Newton-type methods, to name a few. The smart gradient approach can be used to calculate the gradient (and Hessian) numerically, more accurately by using previous descent directions and a transformed coordinate basis. Instead of using the canonical basis at each step, a new orthonormal basis is constructed based on the previous direction using the Modified Gram-Schmidt orthogonalization (see for example \citet{picheny2013}). This transformed basis results in more accurate numeric gradients, which could lead to finding optimums more accurately and more efficiently. For more details see \citet{fattah2022}.
\subsubsection{Expected log-likelihood}
For some likelihoods (as in Section \ref{sec:poisson_example_gen}), the expectation can be calculated analytically, but for others we have to numerically approximate this expectation. Note that as previously stated, the data is assumed conditionally independent given the latent set $\pmb\psi$, and hence the log-likelihood can be constructed by a simple sum of the log-likelihoods from each datapoint. The expected log-likelihood of each datapoint can then be approximated using Gauss-Hermite quadrature, since the integral is with respect to a Gaussian kernel. The expected log-likelihood with respect to the approximate posterior of $\pmb\psi$ is,
\begin{equation}
	E_{\pmb\psi\sim N(\pmb\psi_0 + \pmb Q^{-1}_I\pmb\lambda, \pmb Q_0^{-1})}\left[-\log\pi(\pmb y|\pmb\psi)\right] = \int_{\mathbb{R}^m}-\sum_{i=1}^n\log\pi(y_i|\pmb\psi)\phi(\pmb\psi|\pmb\psi_0 + \pmb Q^{-1}_I\pmb\lambda, \pmb Q_0^{-1})d\pmb\psi.
	\label{eq:ex_loglik}
\end{equation}
Now suppose that the design matrix $\pmb A$ links the data to the parameter $\pmb\psi$, then the linear predictors can be calculated as
\begin{equation}
	\pmb\eta = \pmb A \pmb\psi,
\end{equation}
such that the posterior mean for the $i^{\text{th}}$ linear predictor for $y_i$, $\eta_i$ is $\pmb A_{i.} (\pmb\psi_0 + \pmb Q^{-1}_I\pmb\lambda)$, where $\pmb A_{i.}$ is the $i^{\text{th}}$ row of $\pmb A$. Then,
\begin{equation}
	E_{\pmb\psi\sim N(\pmb\psi_0 + \pmb Q^{-1}_I\pmb\lambda, \pmb Q_0^{-1})}[-\log\pi(\pmb y|\pmb\psi)] = E_{\pmb\eta\sim N(\pmb A(\pmb\psi_0 + \pmb Q^{-1}_I\pmb\lambda), \pmb A\pmb Q_0^{-1}\pmb A^\top)}[-\log\pi(\pmb y|\pmb\eta)],
	\label{eq:ex_loglika}
\end{equation}
since $\pi(\pmb y|\pmb\psi)$ only depends on $\psi$ through $\eta$. Since the data are conditionally independent,
\begin{eqnarray*}
	E_{\pmb\eta\sim N(\pmb A(\pmb\psi_0 + \pmb Q^{-1}_I\pmb\lambda), \pmb A\pmb Q_0^{-1}\pmb A^\top)}\left[-\log\pi(\pmb y|\pmb\eta)\right] &=& 
	E_{\pmb\eta\sim N(\pmb A(\pmb\psi_0 + \pmb Q^{-1}_I\pmb\lambda), \pmb A\pmb Q_0^{-1}\pmb A^\top)}\left[-\sum_{i=1}^n\log\pi(y_i|\pmb\eta)\right]\\
	&=& 
	E_{\pmb\eta\sim N(\pmb A(\pmb\psi_0 + \pmb Q^{-1}_I\pmb\lambda), \pmb A\pmb Q_0^{-1}\pmb A^\top)}\left[-\sum_{i=1}^n\log\pi(y_i|\eta_i)\right] \\
	&=& 
	-\sum_{i=1}^nE_{\eta_i}[\log\pi(y_i|\eta_i)].
	\label{eq:ex_loglikb}
\end{eqnarray*}
The univariate expectations are calculate using Gauss-Hermite quadrature with $m_g$ weights $\pmb w$, and roots $\pmb x^w$, such that,
\begin{equation}
	E_{\pmb\psi\sim N(\pmb\psi_0 + \pmb Q^{-1}_I\pmb\lambda, \pmb Q_0^{-1})}[-\log\pi(\pmb y|\pmb\psi)] \approx 
	\frac{-1}{\sqrt{\pi}}\sum_{r=1}^{m_g} \left[w_r\sum_{i=1}^n\log \pi\left(y_i|\eta_i(x^w_r)\right)\right].
	\label{eq:ex_loglik2}
\end{equation}
To optimize \eqref{eq:vb_lambda} numerically, we expand \eqref{eq:ex_loglik2} around $\pmb\lambda = \pmb 0$ using a second order Taylor series expansion such that,
\begin{equation}
	E_{\pmb\psi\sim N(\pmb\psi_0 + \pmb Q^{-1}_I\pmb\lambda, \pmb Q_0^{-1})}[-\log\pi(\pmb y|\pmb\psi)] \approx 
	\text{constant} + \pmb B^\top \pmb A\pmb Q^{-1}_I\pmb\lambda +
	\frac{1}{2}(\pmb A\pmb Q^{-1}_I\pmb\lambda)^\top\text{diag}(\pmb C)\pmb A \pmb Q^{-1}_I\pmb\lambda,
	\label{eq:ex_loglik3}
\end{equation}
where the $i^{\text{th}}$ entries of $\pmb B$ and $\pmb C$, respectively, are,
\begin{eqnarray*}
	B_i =\sum_{r=1}^{m_g} \frac{w_r x^w_r}{ S_{i}}\log \pi\left( y_i|\eta_i = x^w_r S_i + \pmb A_{i\boldsymbol\cdot} \pmb\psi_0\right) 
\end{eqnarray*}
and
\begin{eqnarray*}
	C_i = \sum_{r=1}^{m_g}\frac{w_r [(x^w_r)^2 - 1]}{S_i^2} \log \pi\left(y_i|\eta_i = x^w_r S_i + 
	\pmb A_{i.} \pmb\psi_0 \right),
\end{eqnarray*}
with $S_{i} = \sqrt{\left(\pmb A^\top  \pmb Q_0^{-1} \pmb A\right)_{ii}}$ .

\section{Illustrative example - low-count Poisson regression
	model}\label{sec:poisson_example_gen}
Here we provide the details for a generalized linear model for count
data, where we use a Poisson response model.  

Suppose we have data
$\pmb y$, of size $n$ with covariates $\pmb X$ and random effect covariates $\pmb u$, then
\begin{eqnarray*}
	Y_i|\beta_0,\pmb\beta,\pmb f &\sim & \text{Poisson}(\exp(\eta_i))\\
	\eta_i &=& \beta_0 +\pmb X_i \pmb{\beta}  + \sum_{k=1}^K f^k(\pmb u_k).
\end{eqnarray*} 

\subsection {Expected log-likelihood}

For the Poisson likelihood, we can obtain a closed-form expression of
the expected log-likelihood and no numerical integration is required. Note that
\begin{eqnarray*}
	E_{\pmb\psi|\pmb \theta\sim N(\pmb\psi_1, \pmb Q_0^{-1})}
	\left[-\log \pi(\pmb\psi|\pmb{y})\right] 
	&=& \int_{\mathbb{R}^m} -\log \pi(\pmb\psi|\pmb{y}) \phi(\pmb\psi|\pmb\psi_1, 
	\pmb Q_0^{-1})d\pmb\psi\notag\\
	&=& \int_{\mathbb{R}^m} \sum_{i=1}^n \left(\exp(\pmb A_{i.}\pmb\psi) - \pmb A_{i.}\pmb\psi y_i + \log(y_i!)\right) 
	\phi(\pmb\psi|\pmb\psi_1, 
	\pmb Q_0^{-1})d\pmb\psi\notag\\
	&=& \sum_{i=1}^n\left(\exp\left(\pmb A_{i.}\pmb\psi_{1}+ \frac{(\pmb A^\top \pmb Q_0^{-1}\pmb A)_{ii} }{2}\right) 
	- y_i(\pmb A_{i.}\pmb\psi_{1}) + \log(y_i!)\right).\notag\\      
\end{eqnarray*}
Now, from
\eqref{eq:vb_lambda}, we find $\pmb\lambda$, where
\begin{eqnarray*}
	\tilde{\pmb{\lambda}} = \arg \min_{\pmb\lambda} &&\left[ \sum_{i=1}^n\left(\exp\left(\pmb A_{i.}(\pmb\psi_{0} + \pmb Q_I^{-1}\pmb\lambda)+ \frac{(\pmb A^\top \pmb Q_0^{-1}\pmb A)_{ii} }{2}\right) 
	- y_i(\pmb A_{i.}(\pmb\psi_{0} + \pmb Q_I^{-1}\pmb\lambda)) \right) \right. \notag\\
	&&\left. +\text{KLD}\left(\phi(\pmb\psi|\pmb\psi_0 + \pmb Q^{-1}_I\pmb\lambda,\pmb Q_0^{-1})||\pi(\pmb\psi)\right) \right].
\end{eqnarray*}

\subsection{Simulation results}
In this section we present an example of the proposed method. We
focus on count data with low counts since this is usually a
challenging case because the likelihood is maximized at $-\infty$, and the second-order expansion of the log-likelihood is less accurate. We use MCMC, a Gibbs sampler (using the \emph{runjags} library) and HMC (using Stan) with a burn-in of $10^2$ and a sample of size $10^5$, 
as the gold standard, and compare VBC to the Laplace method in terms of computational efficiency and accuracy. 

We consider the following over-dispersed count model defined as follows for a dataset of size $n$:
\begin{equation}
	y_i\sim \text{Poisson}(\exp(\eta_i)), \quad \eta_i = \beta_0 + \beta_1x_i + u_i,\label{ex1_eq}
\end{equation}
with a sum-to-zero constraint on $\pmb u$, to ensure identifiability of $\beta_0$.
We use $\beta_0 = -1, \beta_1 = -0.5$ and a continuous covariate $x$, simulated as $x\sim N(0,1)$. The overdispersion is simulated as $u_i\sim N(0,0.25)$. We design the study with the intent of having mostly low
counts. We want to perform full Bayesian inference for the latent field 
$\pmb\psi = \{\beta_0, \beta_1, \pmb u\}$, and the linear predictors $\pmb\eta = \{\eta_1, \eta_2,...,\eta_{n}\}$. 
We assume the following illustrative priors,
\begin{eqnarray*}
	\beta_0 \sim t(5),\quad\beta_1\sim U(-3,3)\quad \text{and}\quad \pmb u \sim N(\pmb 0, 0.25\pmb I)
\end{eqnarray*} 
i.e. $\beta_0$ follows a Student's t prior with $5$ degrees of freedom, $\beta_1$ follows a uniform distribution in $(-3,3)$ and the random effects are independent and identically distributed with a fixed marginal precision of $4$. The vector of estimable parameters is thus $\pmb\psi = \{ \beta_0, \beta_1, u_1, u_2, ..., u_n\}$ of dimension $n+2$.

To illustrate the effect of the low-rank correction we apply the VBC only to $\beta_0, \beta_1$ and then propagate the induced corrections to the $n$ random intercepts, $\pmb u$ and the $n$ linear predictors, $\pmb\eta$.
We thus perform a two-dimensional optimization instead of an $(n+2)$-dimensional optimization, as would be necessary with other variational Bayes approaches.  

\begin{figure}[h]
	\includegraphics[width = 5cm]{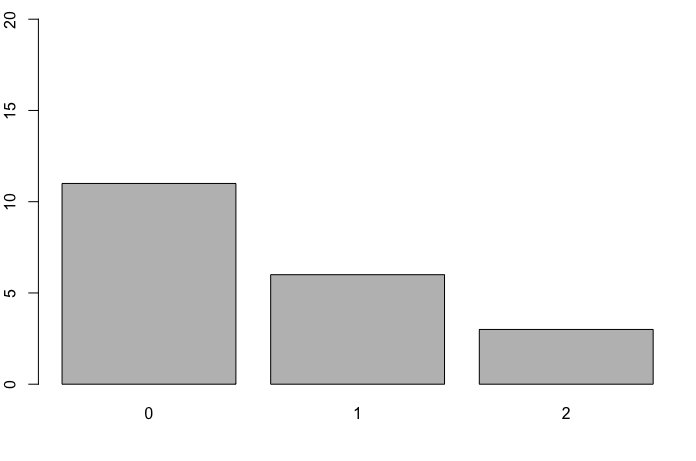}
	\includegraphics[width = 5cm]{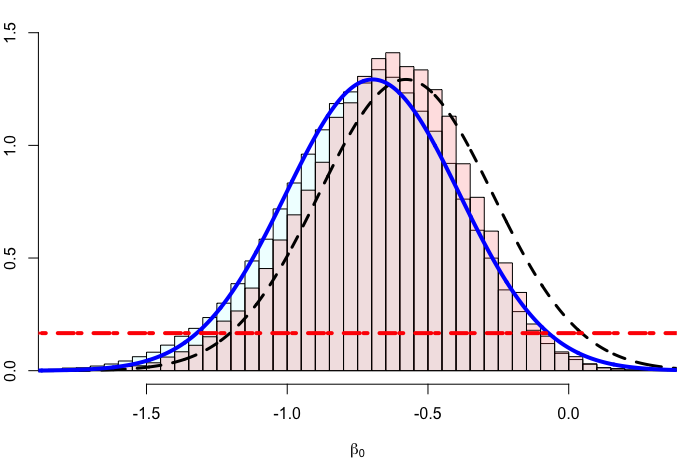}
	\includegraphics[width = 5cm]{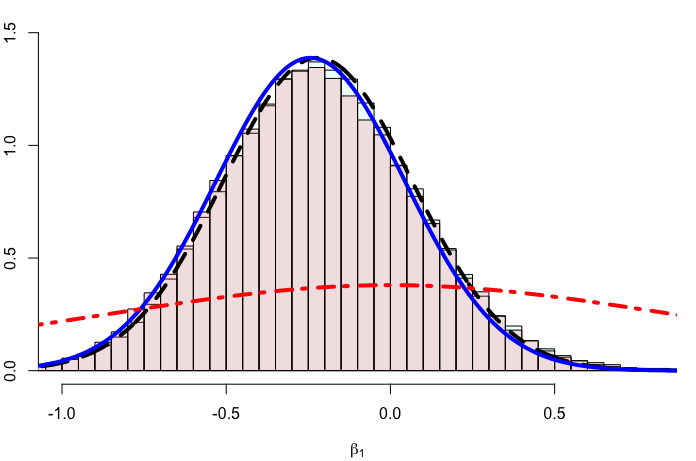}\\
	\includegraphics[width = 5cm]{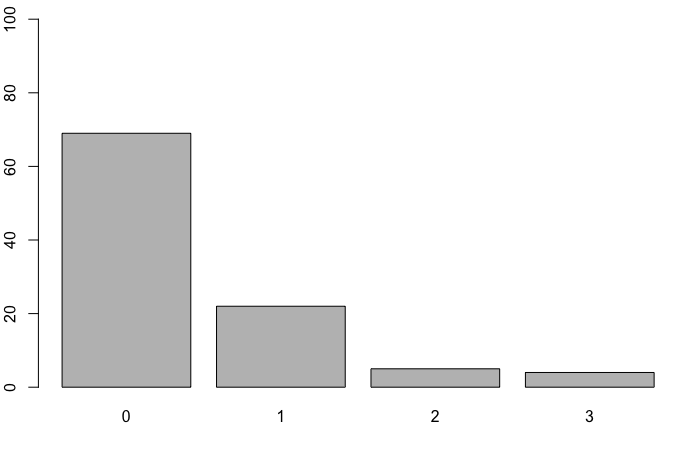}
	\includegraphics[width = 5cm]{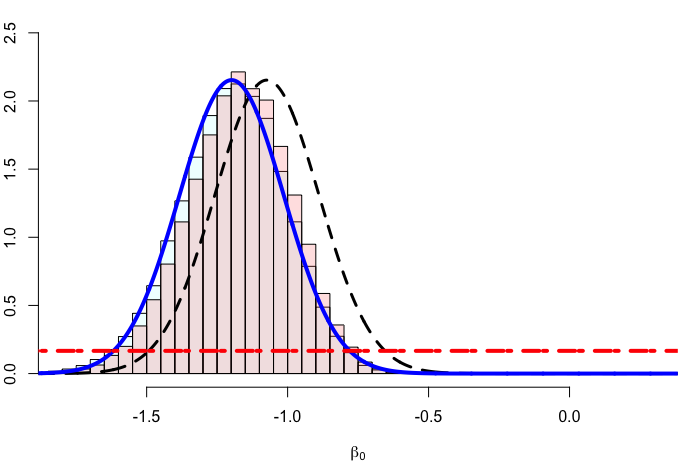}
	\includegraphics[width = 5cm]{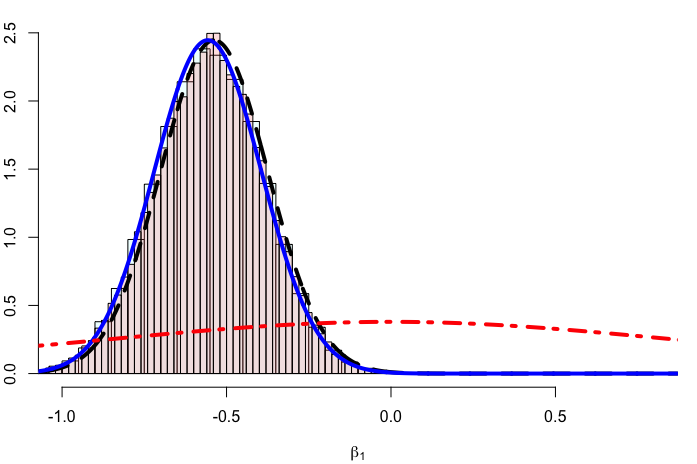}
	\caption{Poisson counts simulated from \eqref{ex1_eq} (left) and
		the marginal posterior of $\beta_0$ (center) and $\beta_1$
		(right) from MCMC (blue histogram), HMC (red histogram), the Laplace method (dashed line) and VBC (solid line) based on the prior (broken line) for $n=20$ (top) and $n=100$ (bottom)}
	\label{fig:sim1data}
\end{figure}

We simulate two samples from the proposed model, one of size $n=20$ and another of size $n=100$, and the data are presented in Figure \ref{fig:sim1data}
(left). 
The posterior means
for the Laplace method, MCMC, HMC and the VBC methods are
presented in Table \ref{table:sim1} for the latent field and selected linear predictors. We can clearly see the improved
accuracy in the mean of the VBC to the Laplace method when compared with the MCMC and HMC output,
from Table \ref{table:sim1} and Figure \ref{fig:sim1data} (center and
right), especially for the smaller dataset. In the case of a larger dataset, we note that the Gaussian approximation performs well and the VBC applies only a slight correction. With the VBC we can achieve similar posterior inference for the latent field and linear predictors, to the MCMC and HMC approaches, more efficiently.
Note that for a small dataset the computational time is small for all the methods, as expected due to the small dimension of
the latent field
$\pmb\psi = \{\beta_0, \beta_1, u_1,u_2, ..., u_{20}\}$ and linear predictors $\pmb\eta = \{\eta_1, \eta_2,...,\eta_{20}\}$,  although for a larger dataset like $n=100$ the excessive computational time for MCMC and HMC is clear, even with this simple model, because the parameter space is of dimension $102$. With the VBC, the correction space is of dimension $2$ and thus the cost of VBC compared to any other inferential framework based on the entire parameter space of dimension $102$, will be much less. The time comparison can be misleading since neither the VBC, MCMC or HMC code has been optimized for this specific model and priors. Nonetheless, the VBC scales well with increasing data size as shown in this example, since the size of the correction space for the optimization stays $2$, while the parameter space size grows from $22$ to $102$. This fictitious example illustrates the stability and scalability of the VBC, for a fixed hyperparameter, which is an unrealistic scenario. In the next Section we provide a hybrid method using the VBC and the integrated nested Laplace approximation (INLA) methodology to address Bayesian inference for more realistic models, including those with hyperparameters, in order to perform Bayesian inference for the latent field and hyperparameters simultaneously.

\begin{table}[h]
	\centering
	\begin{tabular}{|l||l|l|l|l||l|l|l|l|}
		\hline
		& \multicolumn{4}{c}{n=20} & \multicolumn{4}{c}{n=100} \\ \hline
		& LM &VBC & MCMC & HMC & LM & VBC & MCMC & HMC  \\ \hline \hline
		$\beta_0$ & $-0.579$ & $-0.706$ & $-0.715$ & $-0.657$  & $-1.073$ & $-1.199$  & $-1.196$ & $-1.180$ \\ \hline
		$\beta_1$ & $-0.222$ & $-0.224$ & $-0.218$  & $-0.226$ & $-0.538$ &  $-0.567$ & $-0.552$ & $-0.552$  \\  \hline
		$u_1$ & $-0.158$ & $-0.148$ & $-0.152$ & $-0.159$ & $0.177$ & $0.174$ &  $0.175$ & $0.174$ \\  \hline
		$u_8$ & $0.098$ & $0.098$ & $0.099$ & $0.099$ & $-0.046$ & $-0.052$ & $-0.049$ & $-0.044$ \\  \hline
		$u_{15}$ & $ 0.122$ & $0.115$  & $0.118$ & $0.121$ & $-0.074$ & $-0.079$ & $-0.077$ & $-0.073$  \\  \hline
		$\text{Time(s)}$& $2.21$ & $5.78$ & $22.537$ & $12.438$ & $9.48$ & $17.36$ & $384.12$ & $169.57$ \\  \hline
	\end{tabular}
	\caption{Posterior means from the Laplace method, VBC, 
		MCMC and HMC}
	\label{table:sim1}
\end{table}

\section{Application to latent Gaussian models}\label{sec:app_lgm}
The proposal in Section \ref{sec:proposal} can be used to accurately and efficiently calculate the joint posterior for the latent field, only. Many problems, however, include hyperparameters and as such we can embed our proposal into another framework to perform full Bayesian inference for the latent field \emph{and} the hyperparameter set. We use the INLA methodology as proposed by \citet{rue2009} and propose an INLA-VBC methodology. Various strategies, with varying accuracy and efficiency, are available in the INLA framework. The most accurate strategy is the Laplace strategy which involves nested use of the Laplace method, while the least accurate strategy is the Gaussian strategy where the Laplace method is used only once. Naturally, the Gaussian strategy is most efficient while the Laplace strategy is least efficient. Details of these two strategies are presented in the next section. We aim to achieve accuracy similar to that of the Laplace strategy, with a similar cost than that of the Gaussian strategy. We focus our attention to latent Gaussian models (LGMs) for which INLA is developed and show how our proposal can be used.  

A latent Gaussian model appears naturally in statistical modeling since it is a hierarchical Bayesian model with a Gaussian Markov random field as the latent prior. We define an LGM based on data $\pmb{y}$ of size $n$, a latent field $\pmb\psi$ of size $m$ and $n$ linear predictors 
\begin{equation}
	\pmb\eta= \pmb{1}\beta_0 +\pmb X\pmb{\beta}  + \sum_{k=1}^K f^k(\pmb u_k),
	\label{linpred}
\end{equation}
such that $\{\pmb f\}$ are unknown functions (random effects) of $\pmb u$, and
$\pmb \beta$ contains the coefficients for the linear effects of $\pmb X$ on
$\pmb\eta$. The latent field is defined as $\pmb\psi = \{\beta_0, \pmb\beta, \pmb f\}$.
Often, hyperparameters either from the likelihood or the prior of the latent field, form part of the LGM and we denote these by $\pmb\theta$ and assume a moderate dimension $q$ (usually $q<30$). In an LGM the latent field is assumed to follow a Gaussian prior with a sparse precision matrix. The sparseness is often satisfied as most generalized additive mixed models exhibit a sparse precision matrix by construction. Thus an LGM can be summarized as follows:
\begin{eqnarray}
	\pmb y |\pmb\psi, \pmb\theta_1 &\sim& \prod_{i=1}^n\pi(y_i|\pmb\psi, \pmb\theta_1)\notag\\
	\pmb\psi|\pmb\theta_2 &\sim& N(\pmb 0, \pmb Q_\pi^{-1}(\pmb\theta_2))\notag\\
	\pmb\theta = \{\pmb\theta_1, \pmb\theta_2\}&\sim& \pi(\pmb\theta).\label{eq:lgm}
\end{eqnarray}
The aim is thus to estimate the latent posteriors $\pi(\psi_j|\pmb y), j = 1, 2, ...,m$, and the hyperparameter posteriors $\pi(\theta_k|\pmb y), k = 1, 2, ..., q$.  

A specialized methodology called the integrated nested Laplace approximation (INLA) was introduced by \citet{rue2009}, to accurately and efficiently approximate the marginal posteriors of $\pmb\psi$ and $\pmb\theta$ for an LGM. This methodology is based on a series of Gaussian approximations to \emph{conditional} posteriors, using the Laplace method. There is no parametric assumption on the form of the marginal posteriors.
\subsection{INLA}\label{sec:inla}
The INLA methodology from \citet{rue2009} can be summarized as follows,
\begin{eqnarray}
	\pi(\pmb\psi,\pmb\theta, \pmb y) &=& \pi(\pmb\theta)\pi(\pmb\psi|\pmb\theta_2)
	\prod_{i=1}^n\pi(y_i|\pmb\psi,\pmb\theta_1)\notag\\
	\tilde{\pi}(\pmb\theta|\pmb y) &\propto&\left. \frac{\pi(\pmb\psi,\pmb\theta, \pmb y) }
	{\pi_{\text{LM}}(\pmb\psi|\pmb\theta, \pmb y) }\right|_{\pmb\psi = \pmb\psi_0(\pmb\theta)}\notag\\
	\tilde{\pi}(\theta_j|\pmb y) &=& \int \tilde{\pi}(\pmb\theta|\pmb y)d\pmb\theta_{-j}\notag\\
	\tilde{\pi}(\psi_j|\pmb y) &=& \int \tilde{\pi}(\psi_j|\pmb{\theta},\pmb y)
	\tilde{\pi}(\pmb\theta|\pmb y)d\pmb\theta\label{eq:inla},
\end{eqnarray}
where $\pi_{\text{LM}}(\pmb\psi|\pmb\theta, \pmb y)$ is the approximation based on the Laplace method at the mode $\pmb\psi_0(\pmb\theta)$ with precision matrix $\pmb Q_0(\pmb\theta)$ from \eqref{eq:qmu=b}, and $\tilde{f}(.)$ denotes an approximation to $f(.)$. Note that the Laplace method is used for the approximation based on a fixed $\pmb\theta$. For convenience we will use $\pmb\psi_0$ and $\pmb Q_0$, to denote $\pmb\psi_0(\pmb\theta)$ and $\pmb Q_0(\pmb\theta)$, respectively.  

The approximate conditional posterior of $\psi_j$, $\tilde{\pi}(\psi_j|\pmb{\theta},\pmb y)$, can be calculated in one of two ways, extracted from $\pi_{\text{LM}}(\pmb\psi|\pmb\theta, \pmb y)$ (Gaussian strategy) as
\begin{equation}
	\tilde{\pi}(\psi_j|\pmb{\theta},\pmb y)  \approx \phi(\psi_j|\psi_{0j}, Q_0^{jj} ),
	\label{eq:lm_inla}
\end{equation}
or subsequent Gaussian approximations (Laplace strategy) as follows,
\begin{equation}
	\tilde{\pi}(\psi_j|\pmb{\theta},\pmb y) \propto \left. \frac{
		\tilde{\pi}(\pmb\psi,\pmb{\theta}|\pmb y)}{\pi_{\text{LM}}(\pmb\psi_{-j}|\psi_j,\pmb\theta, \pmb y) }\right|_{\pmb\psi_{-j} = \pmb\mu_{-j}(\pmb\theta)},\label{eq:inla_condx}
\end{equation}
where $\pmb\mu_{-j}(\pmb\theta)$ is the mode from the Gaussian approximation to $\pi(\pmb\psi_{-j}|\psi_j,\pmb\theta, \pmb y)$ based on the Laplace method, and $\pmb\psi_{-j}$ is $\pmb\psi$ without the $j^\text{th}$ element. As the dimension of the latent field grow it is clear that the Laplace strategy will become costly due to the multiple Gaussian approximations. It was shown by \citet{rue2009} and multiple works there after that the posteriors from INLA using the Laplace strategy is accurate when compared with those obtained from MCMC sampling, while being much more time and memory efficient than MCMC sampling, even for a large hyperparameter set due to parallel integration strategies (\citet{gaedke2023}). The Gaussian strategy is more efficient, but the resulting marginal posteriors are not accurate enough for some cases. Our proposal in Section \ref{sec:proposal} thus fits naturally into this framework, where we can find a more accurate Gaussian approximation based on the Laplace method and the VBC, by correcting the mode of \eqref{eq:lm_inla}. 

\subsection{INLA-VBC}\label{sec:inla_vbc}
The INLA methodology provides a deterministic framework to approximate the posteriors of the hyperparameters as well as the latent field elements. We apply the proposal from Section \ref{sec:proposal} to the INLA methodology with the hope of achieving more efficient yet accurate approximations of the latent posteriors. Conditional on the hyperparameters, $\pmb\theta$,  define the corrected posterior mean of the joint conditional as $\pmb\psi_1 = \pmb\psi_0 + \pmb\delta$,
where we calculate $\pmb\delta$ implicitly from the correction to $\pmb b_0$ such that
$\pmb b_1 = \pmb b_0 + \pmb\lambda$, where $\pmb\lambda$ is non-zero, only for those elements in $I$, the set of $p$ indices to which we formulate the explicit correction. Note that the latent prior $\pi(\pmb\psi|\pmb\theta_2)$ is Gaussian by construction of the LGM as in \eqref{eq:lgm}, so the KLD term simplifies to the KLD between two multivariate Gaussian densities. Then from \eqref{eq:vb_lambda} and \eqref{eq:ex_loglik3}, we solve for $\pmb\lambda$ (conditionally on $\pmb\theta$) as
\begin{eqnarray}
	\tilde{\pmb\lambda} &=& \arg\min_{\pmb\lambda} \left[E_{\pmb\psi|\pmb\theta\sim N(\pmb\psi_0 + \pmb Q^{-1}_I\pmb\lambda, \pmb Q_0^{-1})}[-\log\pi(\pmb y|\pmb\psi)] + \text{KLD}\left(\phi(\pmb\psi|\pmb\psi_0 + \pmb Q^{-1}_I\pmb\lambda,\pmb Q_0^{-1})||\phi(\pmb\psi|\pmb 0,\pmb Q_\pi)\right)\right]\notag\\
	&=&\arg\min_{\pmb\lambda} \left[E_{\pmb\psi|\pmb\theta\sim N(\pmb\psi_0 + \pmb Q^{-1}_I\pmb\lambda, \pmb Q_0^{-1})}[-\log\pi(\pmb y|\pmb\psi)] + \frac{1}{2}(\pmb\psi_0 + \pmb Q^{-1}_I\pmb\lambda)^\top\pmb Q_{\pi}(\pmb\psi_0 + \pmb Q^{-1}_I\pmb\lambda)\right]
	,\label{eq:vb_lambda_inla}
\end{eqnarray}
where $\pmb Q_I^{-1}$ is constructed from specific columns of $	\pmb Q_0^{-1}$. Thus the improved Gaussian approximation to $\pi(\pmb\psi|\pmb\theta, \pmb y)$ is
\begin{equation}
	\pmb\psi|\pmb\theta, \pmb y\sim N(\pmb\psi_1,  \pmb Q_0^{-1}).
\end{equation}
Now we can use this improved Gaussian approximation to the \emph{conditional} joint posterior, to extract the conditional posteriors for the latent field elements as
\begin{equation}
	\tilde{\pi}(\psi_j|\pmb{\theta},\pmb y) \approx \phi(\psi_j|\psi_{1j},  Q_0^{jj}),\label{eq:vbc_inla}
\end{equation}
instead of the more cumbersome series of Gaussian approximations as in \eqref{eq:inla_condx}. Finally, using the INLA methodology \eqref{eq:inla}, the marginal posteriors of the latent field elements can be calculated as
\begin{equation*}
	\tilde\pi(\psi_j|\pmb y) = \sum_{k=1}^K\tilde\pi(\psi_j|\pmb\theta, \pmb y)\tilde\pi(\pmb\theta_k|\pmb y)\Delta_k,
\end{equation*}
where $\{\pmb\theta_1, \pmb\theta_2, ..., \pmb\theta_K\}$ is a set of values calculated from the joint posterior $\tilde{\pi}(\pmb\theta|\pmb y)$ using a central composite design (CCD) \citep{box1951}, and $\Delta_k$ is the step size.  
Thus, the proposed INLA-VBC methodology can be summarized as
\begin{eqnarray}
	\pi(\pmb\psi,\pmb\theta, \pmb y) &=& \pi(\pmb\theta)\pi(\pmb\psi|\pmb\theta_2)
	\prod_{i=1}^n\pi(y_i|\pmb\psi,\pmb\theta_1)\notag\\
	\tilde{\pi}(\pmb\theta|\pmb y) &\propto&\left. \frac{\pi(\pmb\psi,\pmb\theta, \pmb y) }
	{\pi_{\text{LM}}(\pmb\psi|\pmb\theta, \pmb y) }\right|_{\pmb\psi = \pmb\psi_0}\notag\\
	\tilde{\pi}(\theta_j|\pmb y) &=& \int \tilde{\pi}(\pmb\theta|\pmb y)d\pmb\theta_{-j}\notag\\
	\tilde{\pi}(\psi_j|\pmb y) &=& \int \tilde{\pi}_{\text{VBC}}(\psi_j|\pmb{\theta},\pmb y)
	\tilde{\pi}(\pmb\theta|\pmb y)d\pmb\theta\label{eq:inla2},
\end{eqnarray}
where $\tilde\pi_{\text{VBC}}(\psi_j|\pmb\theta, \pmb y)$ is the VB corrected Gaussian approximation from \eqref{eq:vbc_inla}. Next we show how accurate and efficient this proposal is for approximate Bayesian inference of latent Gaussian models, based on a simulated sample from an overdispersed Poisson model.

\subsection{Simulation results}\label{sec:sim_inla}
We use an overdispersed Poisson regression model with Gaussian priors for the latent field such that
\begin{equation}
	y_i\sim \text{Poisson}(\exp(\eta_i)), \quad \eta_i = \beta_0 + \beta_1x_i + u_i,\label{ex2_eq}
\end{equation}
for $i = 1,2,...,n$, where $u_i|\tau\sim N(0, \tau^{-1})$, $\log\tau\sim \text{loggamma}(1,5e^{-5})$, $\beta_0\sim N(0, 1)$ and $\beta_1\sim N(0,1)$.
The data is simulated based on $\beta_0 = -1, \beta_1 = -0.5, \tau = 1$ and a continuous covariate $x$, simulated as $x\sim N(0,1)$. We want to perform Bayesian inference for the latent field 
$\pmb\psi=\{\beta_0,\beta_1, u_1, u_2, ..., u_n \}$, the linear predictors $\{\eta_1, \eta_2, ..., \eta_n\}$ and the set of hyperparameters $\pmb\theta=\{\tau\}$.  

\begin{figure}[h]
	\includegraphics[width = 7cm, trim = 0cm 0cm 0cm 0cm]{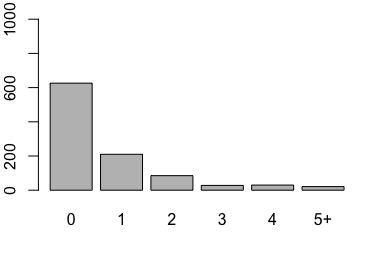}
	\includegraphics[width = 7cm]{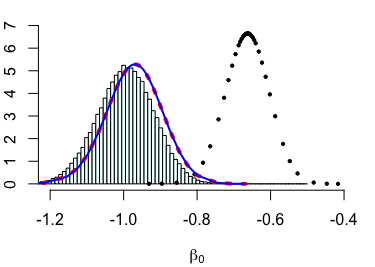}
	\includegraphics[width = 7cm]{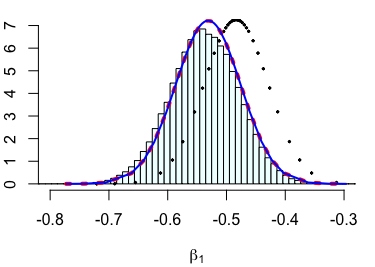}
	\hspace{1cm}
	\includegraphics[width = 7cm]{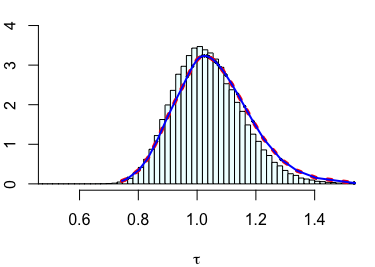}
	\caption{Poisson counts simulated from \eqref{ex1_eq} (top left) and
		the marginal posterior of $\beta_0$ (top right), $\beta_1$
		(bottom left) and $\tau$ (bottom right) from the Gaussian strategy (points), Laplace strategy (dashed line), INLA-VBC (solid line) and MCMC}
	\label{fig:sim1dataA}
\end{figure}

We simulate a sample of $n=1000$ counts 
and the data is presented in Figure \ref{fig:sim1dataA}
(left). In this case we apply the VBC only to the fixed effects $\beta_0$ and $\beta_1$, and the associated changes are then propagated to the posterior means of $\pmb u$ and the linear predictors $\pmb\eta$,
we thus have a two-dimensional optimization instead of a 1002-dimensional optimization as with other variational Bayes approaches, conditional on the hyperparameter $\tau$.  

The posterior means
for the Gaussian strategy (GA), Laplace strategy (INLA), MCMC and the INLA-VBC methods are
presented in Table \ref{table:sim1A}. We can clearly see the improved
accuracy of the INLA-VBC to the Gaussian strategy when compared with the MCMC output,
from Table \ref{table:sim1A} and Figure \ref{fig:sim1dataA} (center and
right), without much additional computational cost based on the time. With the INLA-VBC we can achieve similar results to that of the MCMC approach, more efficiently, while inferring the hyperparameters as well. For the MCMC we used a Gibbs sampler with a burn-in of $10^3$ and a sample of size $10^5$.

\begin{table}[h]
	\centering
	\begin{tabular}{|l||l|l|l|l|l|}
		\hline
		& GA & INLA & INLA-VBC & MCMC  \\ \hline \hline
		$\beta_0$ & $-0.972$ & $-0.664$ & $-0.972$  & $-0.934$ \\ \hline
		$\beta_1$ & $-0.484$ & $-0.532$ &  $-0.531$ & $-0.529$  \\  \hline
		$\tau$ & $1.056$& $1.056$ & $1.056$ & $1.037$ \\ \hline
		$\text{Time(s)}$ & $5.067$ & $18.299$ & $5.718$ & $207.445$  \\  \hline
	\end{tabular}
	\caption{Posterior means from the Gaussian strategy (GA), Laplace strategy (INLA), INLA-VBC and
		MCMC}
	\label{table:sim1A}
\end{table}

\section{Real data examples}\label{sec:real_examples}

We consider two real data examples of different sized datasets. The first example includes a stochastic spline model while the second example is a time to event model based on a continuously-indexed spatial field. Both these models are latent Gaussian models and both involve hyperparameters. We thus use the proposed INLA-VBC methodology from Section \ref{sec:inla_vbc} for full Bayesian inference of these models. Due to the complexity of these models we only compare the results with that of MCMC for the small scale example. Instead, we compare the results based on the Gaussian strategy and the Laplace strategy within the INLA framework of \citet{rue2009}, with those of the INLA-VBC proposal. These examples illustrate the gain in accuracy, without an increased computational cost when using INLA-VBC. 

\subsection{Cyclic second order random walk - small
	scale}\label{sec:tokyorw2}

\begin{figure}
	\includegraphics[width = 5cm]{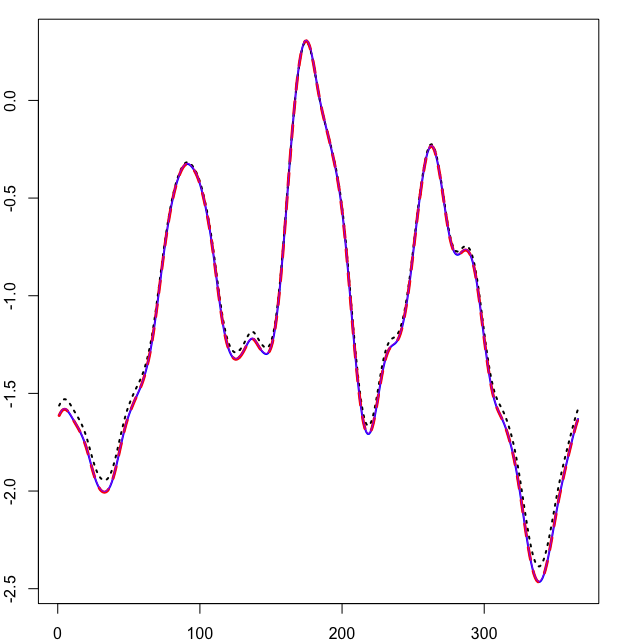}
	\includegraphics[width = 5cm]{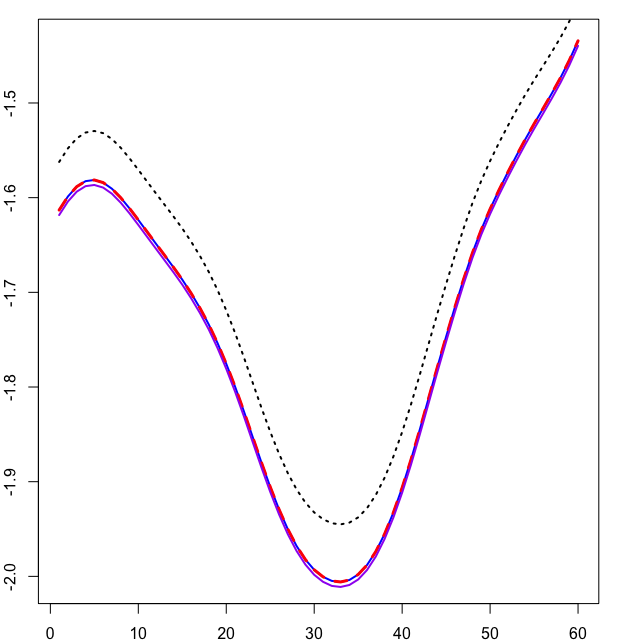}
	\includegraphics[width = 5cm]{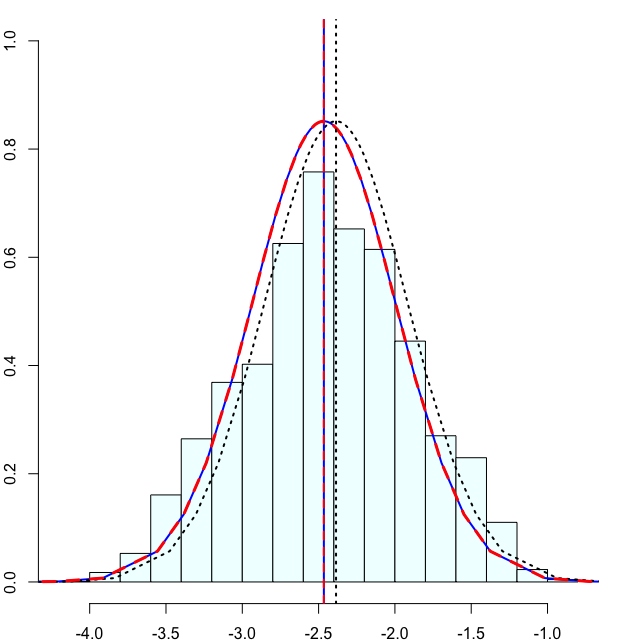}
	\caption{Posterior mean of $\pmb\alpha$ (left) (zoomed for the
		first two months (center)) and the marginal posterior of
		$\alpha_{339}$ (right) from the Gaussian strategy (points), INLA-VBC
		(solid blue line), INLA (broken line) and MCMC (solid purple line and histogram)}
	\label{fig:tokyo}
\end{figure}

The Tokyo dataset \citep{rue2005} in the R-INLA library contains information on the
number of times the daily rainfall measurements in Tokyo was more than
$1$mm on a specific day $t$ for two consecutive years. In order to
model the annual rainfall pattern, a stochastic spline model with fixed precision is used
to smooth the data. In this example we use a cyclic random walk order
two model defined as follows:
\begin{eqnarray}
	y_i|\pmb\psi&\sim& Bin\left(n_i, p_i = \frac{\exp(\alpha_i)}{1+\exp(\alpha_i)}\right)\notag\\
	(\alpha_{i+1}-2\alpha_i+\alpha_{i-1})|\tau&\stackrel{\text{iid}}{\sim}& N(0,\tau^{-1})\notag,
\end{eqnarray}
where $i= 1,2,...,366$, $\pmb\alpha$ is a stochastic spline second order random walk model on a circle (see \citet{rue2005} for more details), and $n_{60} = 1$ else $n_i=2$.
The latent field is $\pmb\psi = \{\pmb\alpha\}$ and we fix the hyperparameter $\tau=1$. Here we apply the correction to $\pmb\alpha$, so that $I = \{1,2,...,n\}$.
In Figure
\ref{fig:tokyo} we present the posterior mean of the spline, estimated with each of the methods
and also the posterior marginal for one specific element, to illustrate the uncertainty in the different posteriors. We can see clearly that the approximate
posterior mean of the spline with INLA-VBC is significantly improved
from the Laplace method, while it is very close to the posterior mean
from INLA and MCMC. All methods estimate similar posterior uncertainty as shown in Figure
\ref{fig:tokyo}.

As a measure of error consolidation, we note that the mean of the
absolute errors produced between the Gaussian strategy and INLA is
$0.0358$ while for INLA-VBC it is $0.0009$, underpinning the findings as
illustrated in Figure \ref{fig:sim1data}. The time for all methods were less than $6$ seconds, as expected due to the small dimension of the data and latent field, although the time for MCMC was $87.63$ seconds. In this real example, the INLA-VBC does not offer much computational gain over the Laplace strategy, although we have a larger difference in computational cost for an increase in data size or model complexity, as shown in the next example.

\subsection{Leukemia dataset - large scale}

\begin{figure}[hbt!]
	\centering \includegraphics[scale=0.5]{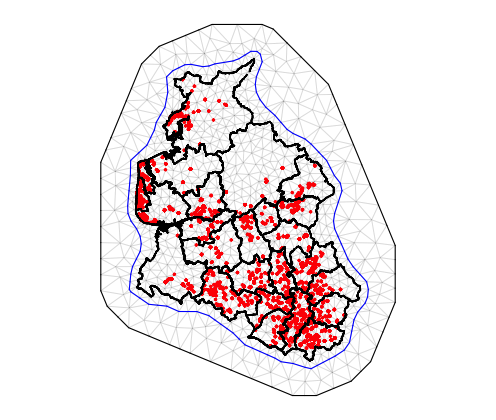}
	\caption{Exact residential locations of patients with AML (dots) and the triangulated mesh for the finite element method estimation of the SPDE of the spatial field}
	\label{fig:amlfig}
\end{figure}

Consider the R dataset \texttt{Leuk} that features the survival times
of $1043$ patients with acute myeloid leukemia (AML) in Northwest
England between 1982 to 1998, for more details see \citet{henderson2002}. 
Exact residential locations and
districts of the patients are known and indicated by the dots in
Figure \ref{fig:amlfig}.
The aim is to model the survival time based on various covariates
$\pmb X$ and space $\pmb s$, with a Cox proportional hazards model,
\begin{center}
	$h(t,\pmb s) = h_0(t) \exp (\pmb\beta\pmb X + \pmb u(\pmb s)),$
\end{center}
where the baseline hazard function $h_0(t)$ is modeled with a stochastic spline as in Section \ref{sec:tokyorw2}, with hyperparameter $\tau$. The baseline
hazard is modeled using $100$ time intervals and we use the data
augmented Poisson regression technique as proposed by
\citet{holford1980analysis, laird1981covariance}. This then implies an
augmented data set of size $11738$.

As fixed effects we use scaled age (\textit{Age}), scaled white blood cell count at
diagnosis (\textit{WBC}) and the scaled Townsend score (\textit{TPI}) as
prognostic factors. Then to account for spatial variation we use a
Gaussian effect $\pmb u$ with a Mat\'ern covariance structure with hyperparameters,
marginal variance $\sigma^2_u$ and nominal range $r = 2/\kappa$ \citep{lindgren2011}. The
model for the linear predictor is
\begin{equation}
	\eta_i(s) = \beta_0 + \beta_1\text{Age}_i + \beta_2\text{WBC}_i + \beta_3\text{TPI}_i + u(s).
\end{equation}
The model for $\pmb{u}$ is continuously indexed in space and we use
the finite element method and the mesh presented in Figure
\ref{fig:amlfig} to estimate this model (for more details regarding
the SPDE approach to continuous spatial modeling see
\citet{lindgren2011,lindgren2015bayesian,krainski2018advanced}).
The mesh contains $2032$ triangles, and through the mapping of the
data to the mesh, we get an augmented latent field of size
$m = 39158$. The hyperparameters to be estimated is
$\pmb\theta = \{\tau, \sigma^2_u, r\}$. Here we apply the correction to the fixed effects only, hence $p=4$, while the other
$m = 39154$ corrections implicitly follow. 
We present the fixed effects posterior means for this example in Table
\ref{table:leuk}, as well as the computational time (which includes the time
to estimate $\pmb\theta$). It is clear that we can achieve the same accuracy as
the Laplace strategy (INLA), at a fraction of the
computational cost with the INLA-VBC.

\begin{table}[h]
	\centering
	\begin{tabular}{|l||l|l|l|l|l|}
		\hline
		& GA & INLA & INLA-VBC   \\ \hline \hline
		$\beta_0$ & $-2.023$ & $-2.189$ & $-2.189$  \\ \hline
		$\beta_1$ & $0.596$ & $0.597$ & $0.597$   \\  \hline
		$\beta_2$ & $0.242$ & $0.241$ & $0.241$  \\ \hline
		$\beta_3$ & $0.108$ & $0.108$ & $0.108$  \\ \hline
		$\tau$ & $0.340$  & $0.340$  & $0.340$ \\ \hline
		$\sigma_u$& $0.223$  & $0.223$  & $0.223$  \\ \hline
		$r$& $0.202$  & $0.202$  & $0.202$  \\ \hline
		Time(s) & $25.9$ & $1276$ & $26.3$  \\ \hline
	\end{tabular}
	\caption{Posterior means from the Gaussian strategy, INLA and INLA-VBC -
		all fixed effects are significant (see Figure \ref{fig:leuk})}
	\label{table:leuk}
\end{table}

\begin{figure}
	\centering \includegraphics[width = 5cm]{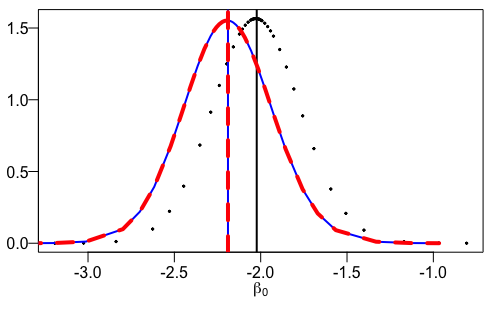}
	\includegraphics[width = 5cm]{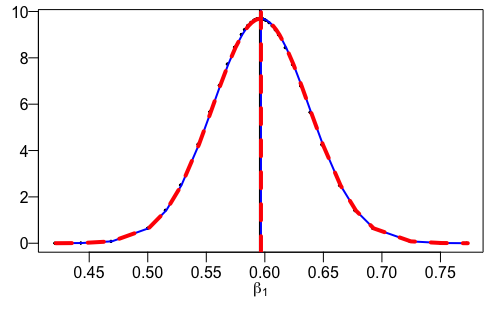}
	\includegraphics[width = 5cm]{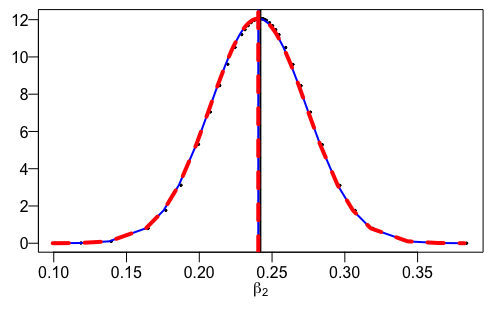}
	\includegraphics[width = 5cm]{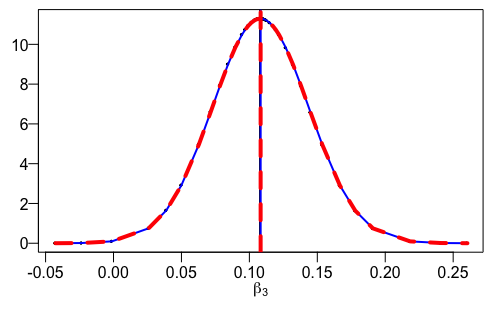}
	\includegraphics[width = 5cm]{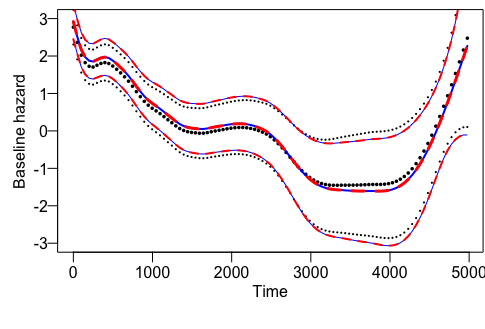}
	\includegraphics[width = 5cm]{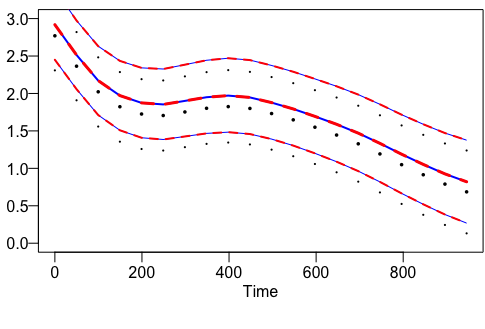}	
	\caption{Marginal posteriors from the Gaussian strategy (points), INLA-VBC
		(solid line) and INLA (broken line) for the fixed effects and posterior mean and $95\%$ credible interval for the baseline hazard $h_0(t)$}
	\label{fig:leuk}
\end{figure}

\begin{figure}
	\centering \includegraphics[width = 5cm]{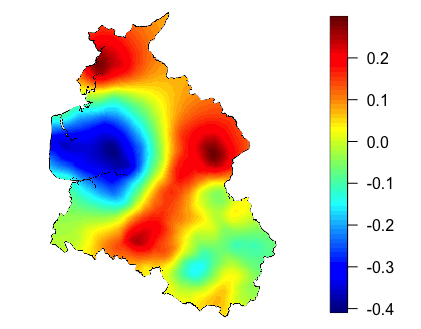}
	\includegraphics[width = 5cm]{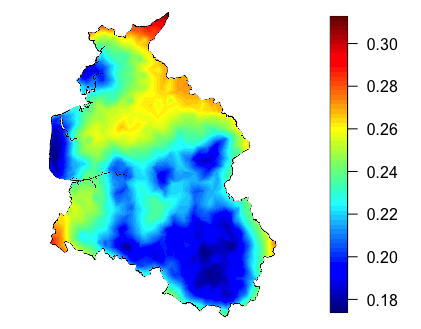}
	\includegraphics[width = 5cm]{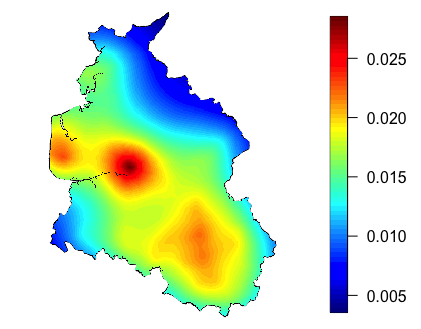}
	\caption{Posterior mean (left) and posterior standard deviation (center) of $\pmb u(\pmb s)$ from INLA-VBC
		with the absolute difference between the posterior means of $\pmb u(\pmb s)$ from the Gaussian strategy and INLA-VBC (right) }
	\label{fig:leuk_gf}
\end{figure}

The marginal posteriors of $\beta_0,\beta_1,\beta_2$ and $\beta_3$ are
presented in Figure \ref{fig:leuk} and the accuracy of the correction
is clear. Note that for $\beta_1, \beta_2$ and $\beta_3$, the posterior means
from the Gaussian strategy are already very close to those from INLA. In
this case we see that the correction from INLA-VBC is stable by estimating only a slight correction. The posterior mean and $95\%$ credible interval of the baseline hazard $h_0(t)$ is presented in Figure  \ref{fig:leuk} and we see that even though we only explicitly correct the four fixed effects, the posterior mean of the baseline hazard is also corrected. Additionally, the posterior mean of the Gaussian field, $\pmb u(\pmb s)$ is presented in Figure \ref{fig:leuk_gf} for INLA-VBC (left), as well as the posterior standard deviation (center). Based on the posterior mean of $\pmb u(\pmb s)$  we can clearly identify areas which have increased risk (red) of death due to AML and also areas where the risk is lower (blue). These areas can be used to inform public health interventions to be targeted towards those areas in need. Since we propagated the explicit correction of the fixed effects to the spatial field as well, we see that corrections were made to the Gaussian field based on the INLA-VBC strategy for most locations (see Figure \ref{fig:leuk_gf} (right)).  

This real example illustrates the potential and need of our proposal, to perform more accurate approximations to the posterior mean (and thus point estimates) for \emph{all} model components (in this example we have made $m = 39158$ improvements to the joint posterior mean) by calculating an optimization in a very small dimension (in this example we solved \eqref{eq:vb_lambda_inla} with $p=4$). 


\section{Discussion and future directions}\label{sec:discussion}

In this paper we proposed a method to correct the posterior mean from a Laplace method using a low-rank correction that propagates to a higher dimensional latent parameter space. We use a variational framework to calculate this lower-dimensional correction. This proposal is useful for problems where a Gaussian approximation from the Laplace method is used to approximate an unknown function, or as an intermediate step in a specific algorithm. We show that the VBC works well compared to MCMC for unimodal unknown functions, in terms of location and uncertainty estimation. Moreover, we apply the VBC to the INLA methodology and construct INLA-VBC that performs full Bayesian inference for latent Gaussian models in an accurate and efficient manner. INLA-VBC achieves similar accuracy in the mean than that of more costly procedures (like MCMC and the Laplace strategy of INLA), without much additional computational cost to that of the Laplace method. INLA-VBC is implemented in the \emph{R-INLA} library and available for use with the \emph{inla} function, more details are available at \url{www.r-inla.org}.  

VBC is not merely a different technique to perform optimization for a Variational Bayes problem, but rather poses a new variational optimization that can be defined on a much smaller dimension than the dimension of the parameter space, while providing results for the entire parameter space. As such, VBC is not to be pinned against other VB computational approaches like inducing point methods, stochastic variational inference, minibatching, boosting approaches, normalizing flow etc, but rather proposes a new framework within which these techniques can be applied.  

VBC can also be used to do a VB
(Bayesian) correction to the maximum likelihood estimator (MLE). This results in an approximate
Bayesian inference framework, starting from a Frequentist basis. The MLE, $\pmb\mu$, of
$\pmb\psi$ is calculated as
\begin{equation*}
	\pmb\mu =      \arg\max_{\pmb\psi} \sum_{i=1}^n\log f(y_i|\pmb\psi).
\end{equation*}
We can also
calculate the precision matrix $\pmb Q$, for $\pmb\psi$
from the Hessian of the log-likelihood at $\pmb\mu$. Asymptotically,
$\pmb\psi\sim N(\pmb\mu, \pmb Q^{-1})$.
Similar to Section \ref{sec:proposal} we impose a low-rank implicit
correction to $\pmb\mu$. We postulate a Gaussian posterior of
$\pmb\psi$ with the corrected mean
\begin{equation*}
	\pmb\mu_1 = \pmb\mu + \pmb Q_I^{-1}\pmb\lambda,
\end{equation*}
and solve for $\pmb\lambda$ using \eqref{eq:vb_lambda}.
This corrected posterior mean then provides
the scientist with a Bayesian estimator adapted from MLE, without
performing a complete Bayesian analysis.  

VBC has the potential to be used also for marginal variance correction and possibly even skewness correction when we move from the Gaussian posterior to a skew-normal posterior family, with a Gaussian copula. As shown in the simulated and real examples, often the variance resulting from the Laplace method is quite accurate when compared with that of MCMC and only a slight correction would be necessary. In the non-latent Gaussian model example the posteriors did not depart significantly from symmetry and the Gaussian posterior appears to be sufficient. In the case latent Gaussian models where we proposed INLA-VBC, the marginal posteriors are not assumed to be symmetric since the integration over the hyperparameter space induces skewness to the Gaussian \emph{conditional} posterior where the VBC was applied, and from the examples considered here the resulting marginal posteriors compare well with that of MCMC. However, scenario's could arise where a variance and skewness correction are beneficial and we are currently exploring these avenues. Initial work in this
area is promising, although the task at hand is more demanding.  

The work we present herein is based on using the variational concept
in an interesting and promising fashion, and we believe that it
contributes to the field of approximate Bayesian inference for a large class of models
as well as
to approximate methods in general by producing accurate results with
superior computational efficiency and scalability.

The examples presented herein can be reproduced based on the code available at \url{https://github.com/JanetVN1201/Code_for_papers/tree/main/Low-rank%20VB%20correction%20to%20GA}.

	\section{Appendix: Optimization-based view of Variational Bayes}
	The Variational Bayes framework as proposed by \citet{zellner1988optimal} can be summarized as follows.  
	
	Based on prior information $\mathcal{I}$, data $\pmb y$ and parameters
	$\pmb{\theta}$, define the following:
	\begin{enumerate}
		\item $\pi(\pmb\theta|\mathcal{I})$ is the prior model assumed for
		$\pmb\theta$ before observing the data
		\item $q(\pmb\theta|\mathcal{D})$ is the learned model from the prior
		information and the data where
		$\mathcal{D} = \{\mathcal{I},\pmb{y}\}$
		\item $l(\pmb\theta|\pmb{y})= f(\pmb{y}|\pmb{\theta})$ is the
		likelihood of state $\pmb\theta$ based on the data $\pmb y$
		\item $p(\pmb{y}|\mathcal{I})$ is the model for the data where
		$p(\pmb{y}|\mathcal{I}) = \int
		f(\pmb{y}|\pmb{\theta})\pi(\pmb\theta|\mathcal{I})d\pmb\theta$
	\end{enumerate}
	The input information in the learning of $\pmb\theta$ is given by
	$\pi(\pmb\theta|\mathcal{I})$ and $l(\pmb\theta|\pmb{y})$. An
	information processing rule (IPR) then delivers
	$q(\pmb\theta|\mathcal{D})$ and $p(\pmb{y}|\mathcal{I})$ as output
	information. A stable and efficient IPR would provide the same amount
	of output information than received through the input information,
	thus being information conservative. Thus, we learn
	$q(\pmb\theta|\mathcal{D})$ such that it minimizes
	\begin{eqnarray}
		&&-\int \left[\log\pi(\pmb\theta|\mathcal{I})+\log l(\pmb\theta|\pmb{y})
		\right]q(\pmb\theta|\mathcal{D})d\pmb\theta + \int \left[\log q(\pmb\theta
		|\mathcal{D})+\log p(\pmb{y}|\mathcal{I})\right]q(\pmb\theta|\mathcal{D})
		d\pmb\theta\notag \\
		&=& -\int \log\pi(\pmb\theta|\mathcal{I})q(\pmb\theta|\mathcal{D})d\pmb
		\theta -\int \log l(\pmb\theta|\pmb{y})q(\pmb\theta|\mathcal{D})d\pmb\theta
		+ \int \log q(\pmb\theta|\mathcal{D})q(\pmb\theta|\mathcal{D})d\pmb\theta
		+\log p(\pmb{y}|\mathcal{I})\notag \\
		&\propto& E_{q(\pmb\theta|\mathcal{D})}\left[-\log l(\pmb\theta|\pmb{y})\right]
		+\int \left[-\log\pi(\pmb\theta|\mathcal{I}) + \log q(\pmb\theta|\mathcal{D})
		\right] q(\pmb\theta|\mathcal{D})d\pmb\theta  \notag \\
		&=&E_{q(\pmb\theta|\mathcal{D})}\left[-\log l(\pmb\theta|\pmb{y})\right] +
		\text{KLD}\left[q(\pmb\theta|\mathcal{D})||\pi(\pmb\theta|\mathcal{I})\right]
		\label{eq:vbdef1}
	\end{eqnarray}
	where
	$\text{KLD}\left[a(x)||b(x)\right] = \int \log \frac{a(x)}{b(x)}
	a(x)dx$ is the Kullback-Leibler divergence measure (or relative
	entropy).
	
	\citet{zellner1988optimal} showed that the learned
	$q(\pmb\theta|\mathcal{D})$ corresponds to the posterior density
	derived from Bayes' theorem, and if we define
	$q(\pmb\theta|\mathcal{D})$ to be the true posterior distribution then
	the IPR in \eqref{eq:vbdef1} is $100\%$ efficient. It is optimal in the
	sense that the amount of the input and output information is as close
	to each other as possible (also the negative entropy of
	$q(\pmb\theta|\mathcal{D})$ is minimized relative to
	$\frac{\pi(\pmb\theta|\mathcal{I})l(\pmb\theta|\pmb{y})}{p(\pmb{y}|\mathcal{I})}$).
	
	Here the Variational concept relates to finding the best candidate
	based on assumptions of the
	analytical form of $q(\pmb\theta|\mathcal{D})$, that minimizes
	\eqref{eq:vbdef1}. This view on variational Bayesian inference is
	beneficial since we do not have to assume that
	$q(\pmb\theta|\mathcal{D})$ is decoupled for $\pmb\theta$, like the mean field
	assumption.

\bibliographystyle{apalike}
\bibliography{BioJ1.bib}

\end{document}